\documentstyle[aps,preprint]{revtex}
\begin{document}
\draft{}

\title{
\begin{flushright}
IM SB RAS NNA 6-96
\end{flushright}
The interference in the reaction $e^+e^-\to\gamma\pi^+\pi^-$ and \\
the search for the decay $\phi\to\gamma f_0\to\gamma\pi^+\pi^-$.
\thanks{This work was partly supported by Russian Fund for Basic Research
, grants 94-02-05 188, 96-02-00 548 and by INTAS-94-3986}}

\author{N.N. Achasov, V.V. Gubin and E.P. Solodov
\thanks{The Budker Institute for Nuclear Physics ,\ \ \
Novosibirsk-90,\ \ \ 630090,\ \ \ Russia}}
\address{Laboratory of Theoretical Physics\\
S.L. Sobolev Institute for Mathematics\\
Novosibirsk-90,\ \ \  630090,\ \ \ Russia
\thanks{E-mail: achasov@math.nsc.ru}}
\date{\today}
\maketitle
\begin{abstract}

We describe the interference between amplitudes
$e^+e^-\to\rho\to\gamma\pi^+\pi^-$  and $e^+e^-\to\phi\to\gamma f_0\to
\gamma\pi^+\pi^-$, where  $f_0$ meson is considered in the frameworks of
the four-quark model and  the model of the scalar $K\bar K$ molecule.
The general expressions for the differential cross section with the
radiative corrections and two angle cuts are given. The interference patterns
are obtained in the spectrum of the differential cross section by the energy
of the photon and in the full cross section by the energy of the beams.

\end{abstract}


\subsection{ Introduction.}
The elucidation of puzzle of scalar  $f_0$ and $a_0$ mesons has became
the central problem of light hadron spectroscopy. As it is known the
properties of the scalar $f_0$ and $a_0$ mesons are mysterious from the
naive quark model point of view.
The long study of these mesons
 \cite{{achdevshest},{achasov-91},{achshest}}  has shown that all challenging
 properties of  the $f_0$ and $a_0$ mesons can be described naturally
in the framework of the four-quark
 $(q^2\bar q^2)$ MIT-bag model \cite{jaffe-77}. Along with it the other
possibilities are discussed in the literature
\cite{{weinstein-90},{tornqvist},{braun},{gribov}}:  the model of scalar
$K\bar K$ molecules, glueballs and so on. This model variety has risen up
the question of looking for the processes permitting to choose the most
adequate one from all abundance.
During years of time there was established by efforts of theorists that the
study of the radiative decays $\phi\to\gamma f_0\to\gamma\pi\pi$ and
$\phi\to\gamma a_0\to\gamma\pi\eta$  could play a crucial role in the
elucidation of the nature of the scalar $f_0$ and  $a_0$ mesons
\cite{{achasov-89},{achasov-95},{close},{molecule}}.

At present time the investigation of the  $\phi\rightarrow\gamma
f_0\rightarrow\gamma\pi^+\pi^-$  decay has started with the detector CMD-2
\cite{novo} at  the $e^+e^-$-collider VEPP-2M in Novosibirsk.
Besides that, in Novosibirsk at the same collider the detector SND has put
into operation \cite{snd} and now it has been working  with
$e^+e^-\to\gamma f_0\rightarrow\gamma\pi^0\pi^0$ and  $e^+e^-\to\gamma a_0
\rightarrow\gamma\eta\pi^0$ decays. The modernization
of the VEPP-2M complex has been planed aiming to increase the luminosity
to one order of magnitude. And,
finally, in the nearest future in Frascati the start of the operation of
the  $\phi$-factory DA$\Phi$NE is expected, which, probably, makes possible
studying the scalar $f_0(980)$ and $a_0(980)$ mesons in exhaustive way.

Experimentally the radiative decays  $\phi\rightarrow\gamma
f_0\rightarrow\gamma\pi\pi$ are studied observing the interference patterns
in the reaction $e^+e^-\to\gamma\pi\pi$ at the $\phi$ meson peak. Analysis of
interference patterns in these reactions, especially
in the charged channel $e^+e^-\to\gamma\pi^+\pi^-$, is the rather difficult
problem to which a great attention was paid in the literature
 \cite{{bramon},{franzini},{lucio}}. But the careful examination of the 
literature has shown that the analysis of interference patterns
in the reaction  $e^+e^-\to\gamma\pi^+\pi^-$ was not only carried out in
exhuastive way but also was improper either from the theoretical point
of view or from the experimental one.
 In particular, the intermediate $K\bar K$ states, at the thresholds
 of which the scalar resonances lie, was not taken into account
in the propagators. In papers \cite{{bramon},{franzini},{lucio}} also
there was not taken into account  the fact that the narrow
width approximation  is not valid in considered case \cite{inudequacy},
so that all branching ratios of the radiative decays of the $\phi$ meson
into the scalar $a_0$ and $f_0$ mesons are at least two times overstated,
 see \cite{inudequacy}. The formulae given in
 \cite{{bramon},{franzini},{lucio}} do not take into account the radiative
corrections that are quite important, see below.

Besides that, all aforementioned papers have studied the interference pattern
in the photon spectrum meanwhile the interference pattern in the full
cross section not only complements that one  but could be  much more
 important in some  particular cases, at low statistics for example.

In this paper we give the full analysis of interference patterns in the
reaction $e^+e^-\to\gamma\pi^+\pi^-$ at the  $\phi$ meson peak
considering two models: the four-quark  $(q^2\bar q^2)$ model and
the model of the scalar  $K\bar K$ molecule. We take our cues from
the preliminary data obtained in the  experiment \cite{novo}.

The paper is organized in the following way.

In the section II we consider the reaction $e^+e^-\to\gamma\pi^+\pi^-$
and give the necessary formulae for this process with taking into account
the cuts of the angle between the photon momentum and electron beam
and of the angle between the photon and $\pi^+$ meson momenta
in the dipion rest frame. We consider the radiative
corrections to the full cross section of the process in that section as well.
The propagators and model depending quantities are described in the
section III. The section IV is devoted to the interference pattern in the
spectrum of the photon energy at the $\phi$ meson peak and the interference
pattern in the full cross section by the total energy of the beams at the
 $\phi$ meson region. In the conclusion we discuss the possibility 
of experimental investigation of interference
patterns in the reaction $e^+e^-\to\gamma\pi^+\pi^-$. The appendix gives
the expressions for the cross section  of the $e^+e^-\to\gamma\mu^+\mu^-$
process which is the background for the $e^+e^-\to\gamma\pi^+\pi^-$ reaction
and gives the values for $BR(\phi\to\gamma^*\to\rho\to\gamma\pi^+\pi^-)$ and
$BR(\phi\to\gamma^*\to\gamma\mu^+\mu^-)$ as well.

\section{Amplitudes
\lowercase{ $e^+e^-\to\phi\to\gamma f_0\to\gamma\pi^+\pi^-$} and
\lowercase{ $e^+e^-\to\rho\to\gamma\pi^+\pi^-$.}}
We consider the production of the $f_0$ meson through the loop of the charged
$K$ mesons, $\phi\to K^+K^-\to\gamma f_0$, see \cite{achasov-89,achasov-95}.
Diagram is presented in Fig. 1a. The production amplitude
 $\phi\to\gamma f_0$  in the rest frame of the $\phi$ meson is:
\begin{equation}
M=g_R(t)\vec e(\phi)\vec e(\gamma)
\end{equation}
where $t=(k_++k_-)^2$, $\vec e(\phi)$ and $\vec e(\gamma)$ are the
polarization vectors of the $\phi$ meson and the photon respectively.
The expressions for $g_R(t)$ are obtained
in the four-quark $(q^2\bar q^2)$ model \cite{achasov-89} and in the
scalar $K\bar K$ molecule model \cite{molecule}. Note, that in the
four-quark model the scalar mesons are considered as a point like objects and
in the  scalar $K\bar K$ molecule model as extended ones \cite{weinstein-90}.

The amplitude of the reaction $e^+e^-\to\phi\to\gamma f_0\to\gamma\pi^+\pi^-$
is
\begin{equation}
M=e\bar u\gamma^{\mu}u\frac{em_{\phi}^2}{f_{\phi}}\frac{g_{f_0\pi\pi}}
{sD_{\phi}(s)D_{f_0}(t)}g_R(t)(q^{\mu}\frac{e(\gamma)p}{pq}-e(\gamma)^{\mu})  \label{amplituda}
\end{equation}
where $s=p^2=(p_1+p_2)^2$, and $g_R(t)\sim(s-t)\sim(pq)\to0$ at
$(pq)\to0\qquad(t\to s)$.
The coupling constants  $g_{f_0\pi\pi}$ and $f_{\phi}$ are related to the
widths in the following way:
\footnote{$\Gamma(f_0\to\pi^+\pi^-,t)=\frac{2}{3}\Gamma(f_0\to\pi\pi,t)$}

\begin{equation}
\Gamma(f_0\to\pi\pi,t)=\frac{g_{f_0\pi\pi}^2\sqrt{t-4m_\pi^2}}{16\pi t},
\qquad\Gamma(V\to e^+e^-,s)=\frac{4\pi\alpha^2}{3}(\frac{m_V^2}{f_V})^2
\frac{1}{s\sqrt{s}}. \label{pipi}
\end{equation}
The width of the $\phi$ meson decay is
\begin{equation}
\Gamma(\phi\to\gamma f_0\to\gamma\pi^+\pi^-)=
\frac{1}{\pi}\int_{4m_{\pi}^2}^{m_{\phi}^2}\sqrt{t}dt\frac{\Gamma(f_0\to
\pi^+\pi^-,t)\Gamma(\phi\to\gamma f_0,t)}{|D_{f_0}(t)|^2},
\end{equation}
where
\begin{equation}
\Gamma(\phi\to\gamma f_0,t)=\frac{1}{3}\frac{|g_R(t)|^2}{4\pi}\frac{1}
{2m_{\phi}}(1-\frac{t}{m_{\phi}^2}).
\end{equation}
The propagators of the $\phi$ and $f_0$  mesons $D_{\phi}(s)$ and
$D_{f_0}(t)$ will be described below.

For the differential cross section we get the expression:
\begin{equation}
\frac{d\sigma_{\phi}(e^+e^-\to\gamma f_0\to\gamma\pi^+\pi^-)}
{dtd\cos\theta_{\gamma}}=\frac{1}{\pi}\frac{\sqrt{t}\Gamma(f_0\to\pi^+\pi^-,t)
\frac{d\sigma}{d\cos\theta_{\gamma}}(e^+e^-\to\gamma f_0,t)}{|D_{f_0}(t)|^2}.
\end{equation}
Having done the integration over angle $\theta_{\gamma}$ we get
\begin{equation}
\frac{d\sigma_{\phi}}{d\omega}=\frac{\alpha^2}{24\pi s^2\sqrt{s}}
\left(\frac{g_{f_0\pi\pi}}{f_{\phi}}\right)^2\frac{m_{\phi}^4}{|D_{\phi}(s)|^2}
\frac{|g_R(t)|^2}{|D_{f_0}(t)|^2}(s-t)\sqrt{1-\frac{\xi}{1-x}}\quad
(a+\frac{a^3}{3})b
\end{equation}
where $\omega=|\vec q|$ is the energy of the photon. Following
 \cite{{bramon},{lucio}} we identify $\xi=\frac{4m_\pi^2}{s}$ and
 $x=\frac{2\omega}{\sqrt{s}}$, $t=s(1-x)$. We also introduce two symmetrical
angle cuts: $-a\leq\cos\theta_{\gamma}\leq a$, where $\theta_{\gamma}$ is
the angle between the photon momentum and the electron beam in the  center
of mass frame of the reaction under consideration  and
 $-b\leq\cos\theta_{\pi\gamma}\leq b$,
where $\theta_{\pi\gamma}$ is the angle between the photon and the $\pi^+$
meson momenta in the dipion rest frame.

As it was shown in the previous papers \cite{{bramon},{franzini}} the basic
background to the process under study has came from the initial
electron radiation  ( see Fig.1d ) and the radiation from the final pions
( Fig.1c ). The initial
state radiation does not interfere with the final state radiation and with
the signal in the differential cross section integrated over all angles
since the charged pions are in the
C=-1 state. This is true also when the angle cuts are symmetrical.

Introducing the symmetrical angle cuts  considerably decreases the
background from the initial state radiation because of the photons in this
case are emitted along the beams mainly. The restriction on the energy of
 photons $20<\omega<100\ MeV$ cuts the background from the radiative
 process with the radiation of hard photons.

In our region $20<\omega<100\ MeV$ the background from the nonresonant by
invariant mass of $\pi^+\pi^-$ system processes, see Fig.1e, is negligible.
Its contribution to
$BR(\phi\to\gamma\pi^+\pi^-,20<\omega<100\ MeV)<2,2\cdot10^{-7}$ and
therefore we do not take into account it.

Let us consider the background related to the final state radiation.
The amplitude of the process is
\begin{eqnarray}
\label{fon}
&&M_{\rho}=e^2\bar u\gamma^{\mu}u\frac{em_{\rho}^2}{f_{\rho}}\frac{1}
{sD_{\rho}(s)}2g_{\rho\pi\pi}T^{\mu}, \\ \nonumber
&&T^{\mu}=\frac{e(\gamma)k_-}{qk_-}(k_+-\frac{p}{2})^{\mu}+
\frac{e(\gamma)k_+}{qk_+}(k_--\frac{p}{2})^{\mu}+e(\gamma)^{\mu}
\end{eqnarray}

It is necessary to take into account the contribution of  the $\phi-\rho$
transition when studying the interference pattern in the full cross section,
see Fig.1c, the quantity  which modulus is as great as  $15\%$
in comparison with the  modulus of the main  term.

Taking into account the vacuum polarization we get
\begin{equation}
M=M_{\rho}(1-Z\frac{m_{\phi}\Gamma_{\phi}}{D_{\phi}(s)})=
M_{\rho}(1-\frac{3\Gamma(\phi\to e^+e^-)\sqrt{s}}{\alpha D_{\phi}(s)}).
\label{firo}
\end{equation}
For simplicity sake, we restricted our expression only to the photon
contribution which is the main in the $\phi-\rho$ transition, see Fig.1f.
Note, that the diagram of Fig.1b is not significant since it is proportional
to  $1/D_{\rho}$ and is negligible in  the $\phi$ meson peak.

It is convenient to give the differential cross section in the form:
\begin{eqnarray}
\label{fonrho}
&&\frac{d\sigma_f}{d\omega}=2\sigma_0(s)\frac{1}{\sqrt{s}}F(x,a,b)|1-
 \frac{3\Gamma(\phi\to e^+e^-)\sqrt{s}}{\alpha D_{\phi}(s)}|^2 \\ \nonumber
&& F(x,a,b)=\frac{2\alpha}{\pi(1-\xi)^{3/2}}\Biggl\{\frac{3}{2}(a-
 \frac{a^3}{3})F_1+\frac{3}{4}a(1-a^2)F_2\Biggr\} \\ \nonumber
&& F_1=\frac{1}{x}\Biggl(x^2-\frac{\xi(1-\xi)(1-x)}{(1-b^2)(1-x)+b^2\xi}
\Biggr)f(x)+
 (1-\xi)(1-x-\frac{\xi}{2})\frac{1}{x}\ln\frac{1+f(x)}{1-f(x)} \\ \nonumber
&& F_2=\frac{1}{x}\Biggl(\frac{\xi^2(x-1)}{(1-b^2)(1-x)+b^2\xi}+2x-2-x^2
\Biggr)f(x)+
 \xi(2-x-\frac{\xi}{2})\frac{1}{x}\ln\frac{1+f(x)}{1-f(x)} \nonumber
\end{eqnarray}
where $f(x)=b\sqrt{1-\frac{\xi}{1-x}}$. The nonradiative cross section
$e^+e^-\to\pi^+\pi^-$ is:
\begin{equation}
\sigma_0(s)=\frac{\pi\alpha^2}{3s}(1-\xi)^{3/2}|F(s)|^2.
\end{equation}
In the vector dominance model the form-factor is
 $|F(s)|^2=(\frac{g_{\rho\pi\pi}}{f_{\rho}})^2\frac{m_{\rho}^4}
{|D_{\rho}(s)|^2}$. We use for the form-factor in  the $\phi$ meson region
the expression:
\begin{equation}
\label{form}
|F(s)|^2=2,6\frac{|D_{\rho}(m_{\phi})|^2}{|D_{\rho}(s)|^2},
\end{equation}
which describes  the experimental data in the $\phi$ meson region
$m_{\rho}^2<s<1,1\ GeV$ \cite{bukin} reasonably good.

The interference between the amplitudes from Eqs. (\ref{amplituda}) and
(\ref{fon}) is equal
\begin{eqnarray}
&&\frac{d\sigma_{int}}{d\omega}=\frac{\alpha^3}{s\sqrt{s}}\left(\frac{
g_{\rho\pi\pi}}{f_{\rho}}\right)\left(\frac{g_{f_0\pi\pi}}{f_{\phi}}\right)
Re\Biggl[\frac{m_{\phi}^2m_{\rho}^2g_R(t)}{\sqrt{4\pi\alpha}D_{\phi}
D^*_{\rho}D_{f_0}}(1-\frac{3\Gamma(\phi\to e^+e^-)\sqrt{s}}{\alpha
D^*_{\phi}(s)})\Biggr]\times \\ \nonumber
&&\times\Biggl\{f(x)+\frac{\xi}{2}\ln\frac{1-f(x)}{1+f(x)}\Biggr\}
(a+\frac{a^3}{3})
\end{eqnarray}

In the similar way let us give the expression  for the differential
 cross section of the initial state radiation.
\begin{eqnarray}
\label{initial}
&&\frac{d\sigma_i}{d\omega}=2\sigma_0(t)\frac{1}{\sqrt{s}}H(x,a,b)|1-
 \frac{3\Gamma(\phi\to e^+e^-)\sqrt{t}}{\alpha D_{\phi}(t)}|^2 \\
&&H(x,a,b)=\frac{\alpha}{\pi}\Biggl[\Biggl(\frac{2(1-x)+x^2}{x}\ln
\frac{1+a}{1-a}-ax\Biggr)(\frac{3b}{2}-\frac{b^3}{2})+
\frac{3a(1-x)(b^3-b)}{x}\Biggr] \nonumber
\end{eqnarray}
Evaluating $H(x,a,b)$ we ignored the electron mass. At $b=1$ our
result coincides with \cite{lucio} ( putting $\beta_e=1$) and differs
by terms of order $x$ with the result quoted by \cite{bramon}
\footnote{See the note in \cite{lucio}.}.

Let us discuss the question about the radiative corrections to the studied
processes. The corrections related to the final
state are proportional to  $L=\ln\frac{s}{m_{\pi}^2}\simeq4$ and
in  the $\phi$ meson peak are small as compared with the initial state
corrections which are proportional to $L=\ln\frac{s}{m_e^2}\simeq16$.
We take into account
the corrections related with the initial state only. If to take into account
that the initial state radiation is approximately twice the final state
radiation  at our angle cuts, see below, then we get that the radiative
corrections to the final state radiation is about $10\%$ of the ones to the
initial state radiation. The general
formulae are obtained in \cite{fadin}. We consider approximate expressions
only.

The total cross section of the one photon annihilation with the soft photon
radiation and with the virtual corrections of order $\alpha$ is given by
\begin{eqnarray}
\label{rad}
&&\sigma(s)=\tilde\sigma(s)\{1+\frac{2\alpha}{\pi}[(L-1)\ln\frac{2
\omega_{min}}{\sqrt{s}}+\frac{3}{4}L+\frac{\pi^2}{6}-1]\}  \\
&&\tilde\sigma(s)=(\sigma_{\phi}(s)+\sigma_{int}(s)+\sigma_{i}+\sigma_{f})
\frac{1}{|1-\Pi(s)|^2}  \nonumber
\end{eqnarray}
where $\omega_{min}$ is the minimal photon energy registered,
$L=\ln\frac{s}{m_e^2}$ is the "main" logarithm.
The given expression is true under condition that $\omega_{min}$
is not larger than the typical resonant width $\Gamma_{res}$. In our case
$\Gamma_{res}\gtrsim25\ MeV$ and  $\omega_{min}=20\ MeV$, so this condition
holds. Much more exact expressions could be found in  \cite{fadin}.
The electron vacuum polarization of order $\alpha$ is
\begin{equation}
\Pi(s)=\frac{\alpha}{3\pi}(L-\frac{5}{3})
\end{equation}
where the contribution of muons and light hadrons is ignored. As one can see
from (\ref{rad}) the radiation corrections lower the cross section by $20\%$.

\section{Production models of \lowercase{$f_0$} meson.}
We consider two models: i) the four-quark $(q^2\bar q^2)$ model and
ii) the model of the scalar $K\bar K$ molecule.

i) In the framework of the four-quark model the  $f_0(980)$ meson is coupled
strongly with the $K\bar K$ channel ( OZI superallowed coupling constant ).
In the paper \cite{achasov-89} the coupling constant of  $f_0$ with
$K^+K^-$ was chosen:
\begin{equation}
\frac{g^2_{f_0K^+K^-}}{4\pi}=2,3\ GeV^2,
\end{equation}
but the other values $g^2_{f_0K^+K^-}/4\pi\simeq1-4\ GeV^2$ are also
acceptable.
The relation $R=g^2_{f_0K^+K^-}/g^2_{f_0\pi^+\pi^-}$ is treated like
a parameter of the model. The processes $\pi\pi\to\pi\pi$ and
$\pi\pi\to K\bar K$ permit the wide enough range for R: $R=4-10$.
When $R=8$, $g^2_{f_0K^+K^-}/4\pi=2,3\ GeV^2$ we get $BR(\phi\to\gamma f_0\to
\gamma\pi\pi)=2,3\cdot10^{-4}$ and effective ( visible) width
$\Gamma_{f_0}\simeq25\ MeV$ \cite{achasov-89}.

In view of the strong coupling constant of the $f_0$ meson with the
$K\bar K$ channel and the vicinity to the $K\bar K$
 threshold it is necessary to take into
account the finite width corrections in the propagator of the $f_0$ meson.
Note, that the finite width corrections distort crucially the ordinary 
resonant Breit-Wigner formulae.

In the four-quark model we treat the propagator in following manner:
\begin{equation}
D_{f_0}(t)=m_{f_0}^2-t+Re\Pi_{f_0}(m_{f_0}^2)-\Pi_{f_0}(t), \label{dqq1}
\end{equation}
where the term $Re\Pi_{f_0}(m_{f_0}^2)-\Pi_{f_0}(t)$ takes into account
the finite width corrections \cite{achdevshest,achasov-95}
\begin{eqnarray}
&&\Pi_{f_0}(t)=\sum_{ab}\Pi^{ab}_{f_0}(t), \nonumber \\
&&Im \Pi^{ab}_{f_0}(t)=\sqrt{t}\Gamma(f_0\to ab,t)=\frac{g_{f_0ab}^2}{16\pi}
\rho_{ab}(t), \nonumber \\
&&\rho_{ab}(t)=\sqrt{\Biggl(1-\frac{m_+^2}{t}\Biggr)\Biggl(1-
\frac{m_-^2}{t}\Biggr)}, \qquad m_{\pm}=m_a\pm m_b.
\end{eqnarray}
The final particle identity is taken into account in the definition of
$g_{f_0 aa}$.

Let $m_a<m_b$, then for $t>m_+^2$
\begin{eqnarray}
&&\Pi^{ab}_{f_0}(t)=\frac{g_{f_0ab}^2}{16\pi}\Biggl[L+\frac{1}{\pi}\rho_{ab}
(t)\ln\frac{\sqrt{t-m_-^2}-\sqrt{t-m_+^2}}{\sqrt{t-m_-^2}+\sqrt{t-m_+^2}}
\Biggr]+i\sqrt{t}\Gamma(f_0\to ab,t), \nonumber \\
&& L=\frac{m_+m_-}{\pi t}\ln(m_b/m_a).
\end{eqnarray}
For $m_-^2<t<m_+^2$ \footnote{ In paper \cite{achasov-95} in (25)
there is a misprint, the third term should have the positive sign,
see ($\ref{misprint}$).}
\begin{equation}
\label{misprint}
\Pi^{ab}_{f_0}(t)=\frac{g_{f_0ab}^2}{16\pi}\Biggl[L-|\rho_{ab}(t)|+
\frac{2}{\pi}|\rho_{ab}(t)|\arctan\frac{\sqrt{m_+^2-t}}{\sqrt{t-m_-^2}}\Biggr].
\end{equation}
For $t<m_-^2$
\begin{equation}
\Pi^{ab}_{f_0}(t)=\frac{g_{f_0ab}^2}{16\pi}\Biggl[L-\frac{1}{\pi}\rho(t)_{ab}
\ln\frac{\sqrt{m_+^2-t}-\sqrt{m_-^2-t}}{\sqrt{m_+^2-t}+\sqrt{m_-^2-t}}\Biggr].
\label{dqq2}
\end{equation}

We consider the finite width corrections for the $f_0$ meson due to the
 $\pi\pi,\ K^+K^-,K^0\bar K^0,\ \eta\eta$ channels as in \cite{achasov-89}.

The calculation of the production amplitude $\phi\to\gamma f_0$ in the
framework of the four-quark model gives the following expression for $g_R(t)$
 \cite{achasov-89}:
when $t<4m_{K^+}^2$
\begin{eqnarray}
&&g_R(t)=\frac{e}{2(2\pi)^2}g_{f_0K^+K^-}g_{\phi K^+K^-}\Biggl\{
1+\frac{1-\rho^2(t)}{\rho(m^2_{\phi})^2-\rho(t)^2}\Biggl[2|\rho(t)|
\arctan\frac{1}{|\rho(t)|}- \nonumber \\
&&-\rho(m^2_{\phi})\lambda(m^2_{\phi})+
i\pi\rho(m^2_{\phi})-(1-\rho^2(m^2_{\phi}))\Biggl(\frac{1}{4}(\pi+
i\lambda(m^2_{\phi}))^2-\Biggl(\arctan\frac{1}{|\rho(t)|}\Biggr)^2
\Biggr)\Biggr]\Biggr\},
\end{eqnarray}
where
\begin{equation}
\rho(t)=\sqrt{1-\frac{4m_{K^+}^2}{t}}\qquad \lambda(t)=\ln\frac{1+\rho(t)}
{1-\rho(t)}.
\end{equation}

When $t>4m_{K^+}^2$
\begin{eqnarray}
&&g_R(t)=\frac{e}{2(2\pi)^2}g_{f_0K^+K^-}g_{\phi K^+K^-}\Biggl\{
1+\frac{1-\rho^2(t)}{\rho(m^2_{\phi})^2-\rho(t)^2}\Biggl[\rho(t)
(\lambda(t)-i\pi)- \nonumber \\
&&-\rho(m^2_{\phi})(\lambda(m^2_{\phi})-i\pi)
-\frac{1}{4}(1-\rho^2(m^2_{\phi}))\Biggl((\pi+i\lambda(m^2_{\phi}))^2-
(\pi+i\lambda(t))^2\Biggr)\Biggr]\Biggr\}.
\end{eqnarray}
The coupling constant $g_{\phi K^+K^-}$ is related to the width:
\begin{equation}
\Gamma(\phi\to K^+K^-)=\frac{1}{3}\frac{g_{\phi K^+K^-}^2}{16\pi}m_{\phi}
\rho(m_{\phi})^3
\end{equation}

ii) The coupling constant in the model of the scalar $K\bar K$ molecule
\cite{close}:
\begin{equation}
\frac{g^2_{f_0K^+K^-}}{4\pi}=0,6\ GeV^2.
\end{equation}

The coupling of the $f_0$ meson with the $K\bar K$ channel in the model of
the $K\bar K$ molecule is considerably weaker than in the four-quark model.

In view of it, we use in the molecular model the propagator of the $f_0$
meson in the traditional Breit-Wigner form.

If $t>4m_{K^+}^2\ ,\enskip 4m_{K^0}^2$\ ,
\begin{eqnarray}
& & D_{f_0}(t)=M_{f_0}^2-t-i\sqrt{t}(\Gamma_0(t)+\Gamma_{K\bar K}(t))
\nonumber\\
& &\Gamma_{K\bar K}(t)=\frac{g^2_{f_0K^+K^-}}{16\pi}(\sqrt{1-4m^2_{K^+}/t}+
\sqrt{1-4m^2_{K^0}/t})\frac{1}{\sqrt{t}}\ .\label{d1}
\end{eqnarray}
If $4m_{K^+}^2 < t < 4m_{K^0}^2$\ ,
\begin{equation}
D_{f_0}(t)=M_{f_0}^2-t+ \frac{g^2_{f_0K^+K^-}}{16\pi}
\sqrt{4m^2_{K^0}/t-1} - i\frac{g^2_{f_0K^+K^-}}{16\pi}
\sqrt{1-4m^2_{K^+}/t}-i\sqrt{t}\Gamma_0(t)\ .
\end{equation} 
When $4m_{K^+}^2\ ,\enskip 4m_{K^0}^2 > t$\ ,
\begin{equation}
D_{f_0}(t)=M_{f_0}^2-t+ \frac{g^2_{f_0K^+K^-}}{16\pi}
(\sqrt{4m^2_{K^+}/t-1}+ \sqrt{4m^2_{K^0}/t-1})-i\sqrt{t}
\Gamma_0(t), \label{d3}
\end{equation}
where the decay width of the scalar $f_0$ resonance into the $\pi\pi$ channel
$\Gamma_0(t)$  is determined by  Eq. (\ref{pipi}).

As a parameter we use the decay width of resonance
$\Gamma(f_0\to\pi\pi,m_{f_0})=\Gamma_0(m_{f_0}^2)=\Gamma_0$.
For $\Gamma_0=50\ MeV$ the effective (visible) width is $\simeq25\ MeV$
and the branching ratio into the  $K\bar K$ channel is
 $BR(f_0\to K\bar K)\simeq0,35$.
For $\Gamma_0=100\ MeV$ the effective (visible) width is $\simeq75\ MeV$ and
the branching ratio into the $K\bar K$ channel
 is $BR(f_0\to K\bar K)\simeq0,3$.

 Since the scalar resonance lies under the $K\bar K$ threshold, the peak
in the cross section or in the mass spectrum does not coincide with
$M_{f_0}$. It is easy to check using Eqs. (\ref{d1})--(\ref{d3}).
Because of this the mass in the Breit-Wigner formulas should be renormalized:
\begin{eqnarray}
M^2_{f_0}=m^2_{f_0}-\frac{g^2_{f_0K^+K^-}}{16\pi}(\sqrt{4m^2_{K^+}/
m_{f_0}^2-1}+\sqrt{4m^2_{K^0}/m_{f_0}^2-1})\ ,
\end{eqnarray}
where $m^2_{f_0}$ is the physical mass square and $M^2_{f_0}$ is
the bare mass square. So, the physical mass is greater then the bare one.
This fact is particularly important when the coupling of scalar meson
with the $K\bar K$ channel is strong as it is in the four-quark  and
molecular models. But this circumstance was not taken into account
neither  in fitting data nor in theoretical papers with the exception of
\cite{achdevshest,achasov-89,achasov-95,molecule,inudequacy,achasov-88}.

Let us note that Eqs. (\ref{d1})--(\ref{d3}) are true
in the resonance region only. They have wrong analytical properties at
$t=0$, for example. The expressions that are free of this trouble are
given above, see Eqs. (\ref{dqq1})--(\ref{dqq2}).

When the scalar resonance lies between the $K\bar K$ thresholds the
renormalization of mass should be done in the following way:
\begin{eqnarray}
M^2_{f_0}=m^2_{f_0}-\frac{g^2_{f_0K^+K^-}}{16\pi}\sqrt{4m^2_{K^0}/
m_{f_0}^2-1}\ .
\end{eqnarray}
Note, that in the molecular model $m_{f_0}-M_{f_0}=24(10)\ MeV$ for
$m_{f_0}=980(2m_{K^+})\ MeV$.

The calculation of the amplitude in the model of the scalar
 $K\bar K$ molecule
was performed in \cite{molecule}. As the analysis of model has shown the
imaginary part of the  production amplitude  $\phi\to\gamma f_0$ gives about
$90\%$ of all intensity of the decay   $\phi\to\gamma f_0\to\gamma\pi\pi$.
Therefore in the model of the scalar $K\bar K$ molecule we consider
the imaginary part of $g_R(t)$ only.

When  $t<4m_{K^+}^2$
\begin{eqnarray}
&&Img_R(t)=\pi eg_{f_0K^+K^-}g_{\phi K^+K^-}\frac{\mu^4}{(2\pi)^2}\frac{1}
{(t-4a^2)^2}\Biggl\{\frac{m_{\phi}^2}{\omega^3}\Biggl(\ln\frac{(E_1-a)
(E_2+a)}{(E_2-a)(E_1+a)}\times\nonumber\\
&\times&\frac{E_1E_2t(12a^2-t)-a^2m_{\phi}^2(t+4a^2)}{4a^3t}\Biggr)+
\frac{4m_{K^+}^2}{\sqrt{t}\omega}\ln\frac{E_1^2-a^2}{E_2^2-a^2}+
\frac{8m_{K^+}^2}{\omega\sqrt{t}}\lambda(m^2_{\phi})-\nonumber\\
&-&\frac{m_{\phi}^2(t-4a^2)\rho(m^2_{\phi})}{2a^2\omega^2}-
\frac{32m_{\phi}^2\rho(m^2_{\phi})^3(t-4a^2)^2}{3(m_{\phi}^2-4a^2)^3}\Biggr\}
\end{eqnarray} 
where $a^2=m_{K^+}^2-\mu^2\qquad$, $p_0=(m_{\phi}^2+t)/2\sqrt{t}\qquad$,
$\omega=(m_{\phi}^2-t)/2\sqrt{t}\qquad$,
$E_1=\frac{1}{2}(p_0-\omega\rho(m^2_{\phi}))$\ , and
$E_2=\frac{1}{2}(p_0+\omega\rho(m^2_{\phi}))$\ .

  When $t>4m_{K^+}^2$
\begin{eqnarray}
&&Img_R(t)=\pi eg_{f_0K^+K^-}g_{\phi K^+K^-}\frac{\mu^4}{(2\pi)^2}
\frac{1}{(t-4a^2)^2}\Biggl\{\frac{m_{\phi}^2}{\omega^3}
\Biggl(\ln\frac{(E_1-a)(E_2+a)}{(E_2-a)(E_1+a)}\times\nonumber\\
&\times&\frac{E_1E_2t(12a^2-t)-a^2m_{\phi}^2(t+4a^2)}{4a^3t}\Biggr)+
\frac{4m_{K^+}^2}{\omega\sqrt{t}}\ln\frac{E_1^2-a^2}{E_2^2-a^2}+
\frac{8m_{K^+}^2}{\omega\sqrt{t}}\lambda(m^2_{\phi})-\nonumber\\
&-&\frac{m_{\phi}^2(t-4a^2)\rho(m^2_{\phi})}{2a^2\omega^2}-
\frac{32m_{\phi}^2\rho(m^2_{\phi})^3(t-4a^2)^2}{3(m_{\phi}^2-4a^2)^3}
+\frac{4m_{\phi}^2\rho(t)}{\omega\sqrt{t}}-\frac{8m_{K^+}^2}{\omega
\sqrt{t}}\lambda(t)\Biggr\}
\end{eqnarray} 
where $\mu=140\ MeV$ \cite{close}.

For the propagator of the $\phi$ meson we use the expression:
\begin{equation}
D_{\phi}(s)=m_{\phi}^2-s-is\frac{g^2_{\phi K^+K^-}}{48\pi}\Biggl[(1-
\frac{4m_{K^+}^2}{s})^{3/2}+c_1(1-\frac{4m_{K^0}^2}{s})^{3/2}\Biggr]-
ic_2p^3_{\pi\rho}(s)
\end{equation}
where $p^2_{\pi\rho}(s)=\sqrt{s}\rho_{\pi\rho}(s)/2$. Taking into account the
branching ratios of the $\phi$ meson decays and the total normalization
we get $c_1=1,09$ and  $c_2=0,1$.

The propagator of the $\rho$ meson is:
\begin{equation}
D_{\rho}(s)=m_{\rho}^2-s-is\frac{g^2_{\rho\pi\pi}}{48\pi}(1-
\frac{4m_{\pi}^2}{s})^{3/2}
\end{equation}

\section{The interference patterns.}

We consider the following parameters in the four-quark model:
$m_{f_0}=980\ MeV$, $R=8$,  $g^2_{f_0K^+K^-}/4\pi=2,3\ GeV^2$, so that
$BR(\phi\to\gamma f_0\to\gamma\pi\pi)=2,3\cdot10^{-4}$ and the visible width
$\Gamma_{f_0}\simeq25\ MeV$ \cite{achasov-89}.
The interference pattern by the total energy of the beams
in the full cross section of the reaction  $e^+e^-\to\gamma\pi^+\pi^-$,
$\sigma=\sigma_{\phi}\pm\sigma_{int}+\sigma_f+\sigma_i$, at
the  $\phi$ peak is shown in Fig.2.
Guided by \cite{novo}, we choose the angle cuts  $a=0,66$ and
$b=0.955$, which  decrease the initial state radiation by a factor of nine.
But, in despite of the strong suppression, the initial state radiation stays
dominant and is equal about  $\frac{2}{3}$ of total background.
The energy of the photon lies in the  interval $20<\omega<100\ MeV$.

The dotted line and the line 1 apply to the pure background
and to the background with  the $\phi-\rho$ transition
respectively. The lines 2 and 3 show constructive and destructive
interference correspondingly.

As one can see from Eq. (\ref{firo}) the contribution of
the  $\phi-\rho$ transition
to the amplitude is about $15\%$ at $\sqrt{s}=m_{\phi}\pm\Gamma_{\phi}/2$.
But, since the initial state radiation, in which  the $\phi-\rho$ transition
is negligible for the $\sqrt{t}<m_{\phi}-20\ MeV$,
forms the major part of  background the relative
contribution of $\phi-\rho$ transition is smaller in the total pattern
than in the  $e^+e^-\to\pi^+\pi^-$ one and is equal about $4\%$ at
$\sqrt{s}=m_{\phi}\pm\Gamma_{\phi}/2$,  as it is seen
from Fig.2.

The interference pattern in the photon spectrum $d\sigma_{\phi}/d\omega\pm
d\sigma_{int}/d\omega$ at the $\phi$ meson point is shown in Fig.3.

In the model of the scalar  $K\bar K$ molecule we use the  parameter:
$\Gamma(f_0\to\pi\pi,m_{f_0}=980\ MeV)=\Gamma_0=50\ MeV$.
For $m_{f_0}=980\ MeV$, $g^2_{f_0K^+K^-}/4\pi=0,6\ GeV^2$ we have
$BR(\phi\to\gamma f_0\to\gamma\pi\pi)=1,7\cdot10^{-5}$ \cite{molecule}.
The visible width is $\simeq25\ MeV$ and the branching ratio into the 
$K\bar K$ channel is $BR(f_0\to K\bar K)\simeq0,35$.
The angle cuts  are the same. The interference patterns are shown
in Fig.4 and Fig.5.

\section{Conclusion.}

The analysis of the graphs presented shows that the observation of the
interference patterns in the reaction $e^+e^-\to\gamma\pi^+\pi^-$
is quite possible at the building $\phi$-factories in Novosibirsk
and Frascati. In the case of  $q^2\bar q^2$ model the observation is possible
at the detectors CMD-2 and SND at the VEPP-2M collider. Furthermore,
the planed experimental statistics at the $\phi$-factories
will let analyzing interference patterns to decide between two models
of the $f_0$ meson makeup: $q^2\bar q^2$ model and the model
of the $K\bar K$ molecule.

Really, as one can see in Fig.2, in the case of $q^2\bar q^2$ model
the difference between the constructive  interference and the background 
at the $\phi$ meson point is $\simeq0,34\ nb$, and for the destructive
interference this difference is $\simeq0,14\ nb$,  at the total cross section
$e^+e^-\to\gamma\pi^+\pi^-$ is $1,6\ nb$.

As it is seen from Fig.4, for the model of the $K\bar K$ molecule
the difference between the destructive interference and the background equals
approximately to the difference between the background and the constructive
one and is  $\simeq0,07\
nb$ at the $\phi$ meson point.

Besides that, the comparison of two graphs shows that the behavior
of the constructive interference in the  $q^2\bar q^2$ case
and in the case of the
$K\bar K$ molecule differs fundamentally. In the $q^2\bar q^2$ case
the constructive interference has a prominent peak while in the case of
the  $K\bar K$ molecule such a peak is absent. This behavior is easy
to be distinguished experimentally since the difference of the
cross sections is  $0,27\ nb$.

For the  destructive interference the difference between two models is not
so strong.
In despite of the fact that the signal in the model
of the scalar $K\bar K$ molecule is much weaker then in the  $q^2\bar q^2$
model ($BR(\phi\to\gamma f_0(molecule)
\to\gamma\pi\pi)\simeq\frac{1}{10}BR(\phi\to\gamma f_0(q^2\bar q^2)
\to\gamma\pi\pi)$, see also \cite{{achasov-89},{molecule}}, the cross section
difference decreases not so much. The reason is that in the case
of the destructive interference the interference term in the  $q^2\bar q^2$
model is compensated by the modulus square of $f_0$ meson production
amplitude meanwhile in the model of the $K\bar K$ molecule the
interference term is dominant.

The difference between the cross sections of the destructive interference
case in the four-quark model and in the model of the $K\bar K$ molecule 
is about  $0.1\ nb$ at the $\phi$ meson peak.

On the other hand, we have to note a quite week dependence of the 
interference pattern in the total cross section on the parameters of the
models. To illustrate it we show the lines of the destructive interference
for the different parameters of the models in Fig.6.

In the four-quark model for $m_{f_0}=980\ MeV$, $R=8$,
$g^2_{f_0K^+K^-}/4\pi=2,3\ GeV^2$, so that
$BR(\phi\to\gamma f_0\to\gamma\pi\pi)=2.3\cdot10^{-4}$,
$BR(\phi\to\gamma f_0\to\gamma\pi\pi,20<\omega<100\ MeV)=1.13\cdot10^{-4}$,
the visible width $\Gamma_{f_0}\simeq25\ MeV$ the curve is shown as a solid
line.

The dotted line is for $m_{f_0}=980\ MeV$, $R=4$,
$g^2_{f_0K^+K^-}/4\pi=4\ GeV^2$, so that
$BR(\phi\to\gamma f_0\to\gamma\pi\pi)=5\cdot10^{-4}$,
$BR(\phi\to\gamma f_0\to\gamma\pi\pi,20<\omega<100\ MeV)=1.14\cdot10^{-4}$
and the visible width $\Gamma_{f_0}\simeq50\ MeV$ in the four-quark model.

The dashed line for  $m_{f_0}=980\ MeV$, $R=1$,
$g^2_{f_0K^+K^-}/4\pi=0,19\ GeV^2$ that correspond the $s\bar s$ structure
of the $f_0$ meson \cite{achasov-89,achasov-95}. At such parameters
 $BR(\phi\to\gamma f_0\to\gamma\pi\pi)=5\cdot10^{-5}$,
$BR(\phi\to\gamma f_0\to\gamma\pi\pi,20<\omega<100\ MeV)=2.4\cdot10^{-5}$
and the visible width $\Gamma_{f_0}\simeq50\ MeV$.

And also the curve without $f_0$ resonance (the line 2), we have for that
$BR(\phi\to\gamma\to\rho\to\gamma\pi\pi,20<\omega<100\ MeV)=3,5\cdot10^{-6}$,
see appendix.

In despite of the fact that the values of the partial widths vary
by two orders the total interference pattern changes not so dramatically.

In parallel with the interference pattern in the total cross section one
should consider
the interference pattern in the photon spectrum which for the visible width
$\Gamma_{f_0}\simeq25\ MeV$ has a good sensitivity since it is a differential
characteristic, see Fig.3 and Fig.5. The concurrent observation
of two interference patterns, by the energy of beam and by the energy of the
photon, extends the possibility
for analysis and allows  to have  more strong  limits. Along with it,
for the broad $f_0$ meson ( the visible width $\Gamma_{f_0}\simeq50\ MeV$)
the interference pattern in the photon spectrum is less informative.

\section{ Appendix.}

In the case of an experimental difficulties to distinguish  the  charged
pions from muons  the detected events $e^+e^-\to\gamma\pi^+\pi^-$ contain
the part of the events from the reaction
$e^+e^-\to\gamma\mu^+\mu^-$, \cite {novo}.
 In this situation it is important to know
the cross section $\sigma(e^+e^-\to\gamma\mu^+\mu^-)$ and the branching ratio
of the  $\phi$ meson decay into the $\gamma\mu^+\mu^-$. In this appendix
we give the necessary expressions and the values for
$BR(\phi\to\gamma^*\to\gamma\mu^+\mu^-)$ and
$BR(\phi\to\gamma^*\to\rho\to\gamma\pi^+\pi^-)$ .

The cross section of the $e^+e^-\to\gamma\mu^+\mu^-$ process consists of the
initial electron radiation, see Eq. (\ref{initial}):
\begin{equation}
\frac{d\sigma_i(e^+e^-\to\gamma\mu^+\mu^-)}{d\omega}=
2\sigma_0(t)\frac{1}{\sqrt{s}}H(x,a,b)|1-\frac{3\Gamma(\phi\to e^+e^-)
\sqrt{t}}{\alpha D_{\phi}(t)}|^2
\end{equation}
where the cross section of the  $e^+e^-\to\mu^+\mu^-$ reaction
\begin{equation}
\sigma_0(s)=\frac{4\pi\alpha}{3s}\sqrt{1-\xi_{\mu}}(1+\frac{\xi_{\mu}}{2}),
\end{equation}
we use the definitions: $\xi_{\mu}=4\mu^2/s$, and the final muon radiation
\cite{xose}:
\begin{equation}
\frac{d\sigma_f(e^+e^-\to\gamma\mu^+\mu^-)}{d\omega}=
2\sigma_0(s)\frac{1}{\sqrt{s}}F_{\mu}(x,a)|1-\frac{3\Gamma(\phi\to e^+e^-)
\sqrt{s}}{\alpha D_{\phi}(s)}|^2,
\end{equation}
where
\begin{eqnarray}
&&F_{\mu}(x,a)=\frac{3\alpha}{4\pi\sqrt{1-\xi_{\mu}}(1+\frac{\xi_{\mu}}{2})}
\Biggl[-a(x+\frac{1-x}{x}\xi_{\mu})f_{\mu}-\frac{a^3}{3}(x+\frac{1-x}{x}
(8+\xi_{\mu}))f_{\mu}+ \\ \nonumber
&&+a(x+\frac{1-x}{x}(2+\xi_{\mu})-\frac{\xi_{\mu}}{x}(1+\frac{\xi_{\mu}}{2}))
\ln\frac{1+f_{\mu}}{1-f_{\mu}}+\frac{a^3}{3}(x+\frac{1-x}{x}(2+\xi_{\mu})+
\\ \nonumber
&&+\frac{\xi_{\mu}}{x}(1-\frac{\xi_{\mu}}{2}))
\ln\frac{1+f_{\mu}}{1-f_{\mu}}\Biggl].
\end{eqnarray}
and  $f_{\mu}=\sqrt{1-\xi_{\mu}/(1-x)}$, see Fig.1g.

The sum of the initial electron radiation and the final muon radiation  at
$a=0.66$ and $20<\omega<100\ MeV$ is shown in Fig.7. The initial radiation
is about a half of the total background, in view of this the $\phi-\gamma$
contribution is suppressed relatively less than in the
 $e^+e^-\to\gamma\pi^+\pi^-$
process and is equal about $7\%$ at $\sqrt{s}=m_{\phi}\pm\Gamma_{\phi}/2$.

The decay width $\phi\to\gamma^*\to\gamma\mu^+\mu^-$ is given in the
following way:
\begin{equation}
\frac{d}{d\omega}\Gamma(\phi\to\gamma\mu^+\mu^-)=\frac{2}{m_{\phi}}
\Gamma(\phi\to\mu^+\mu^-)F_{\mu}(x)
\end{equation}
where
\begin{equation}
\Gamma(\phi\to\mu^+\mu^-)=\Gamma(\phi\to e^+e^-)(1+\frac{2\mu^2}{m_{\phi}^2})
\sqrt{1-\frac{4mu^2}{m_{\phi}^2}}
\end{equation}
and $F_{\mu}(x)=F_{\mu}(x,a=1)$.

After integrating over photon energy in the range
$20<\omega<100\ (\omega_{max}\simeq470\ MeV)$ we get:
\begin{equation}
BR(\phi\to\gamma\mu^+\mu^-)=7,3\cdot10^{-6}(1,15\cdot10^{-5}).
\end{equation}
Analogously the decay width $\phi\to\gamma^*\to\rho\to\gamma\pi^+\pi^-$
is given in the form:
\begin{equation}
\frac{d}{d\omega}\Gamma(\phi\to\gamma\pi^+\pi^-)=\frac{2}{m_{\phi}}
\Gamma(\phi\to\pi^+\pi^-)F(x)
\end{equation}
where $F(x)=F(x,a=1,b=1)$, see Eq. (\ref{fonrho}), and
\begin{equation}
\Gamma(\phi\to\pi^+\pi^-)=\frac{1}{4}\Gamma(\phi\to e^+e^-)
(1-\frac{4m_{\pi}^2}{m_{\phi}^2})^{3/2}|F(m_{\phi}^2)|^2.
\end{equation}
The form-factor $|F(m_{\phi}^2)|^2=2,6\ $,  see Eq. (\ref{form}).
After integrating in the range $20<\omega<100\ (\omega_{max}\simeq470\ MeV)$
 we get:
\begin{equation}
BR(\phi\to\gamma\pi^+\pi^-)=3,5\cdot10^{-6}(4,7\cdot10^{-6}).
\end{equation}

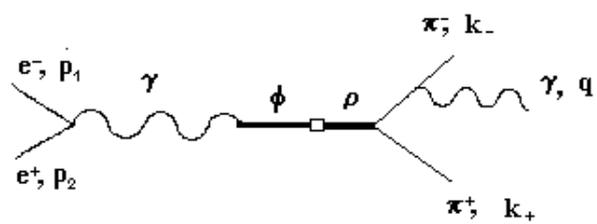
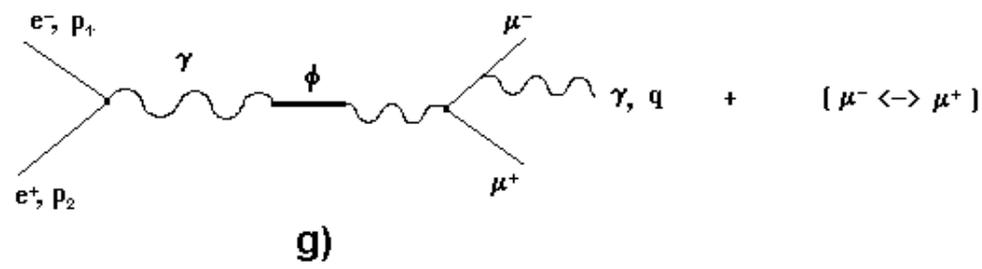
\begin{figure}
\caption{Model diagrams.}
\end{figure}
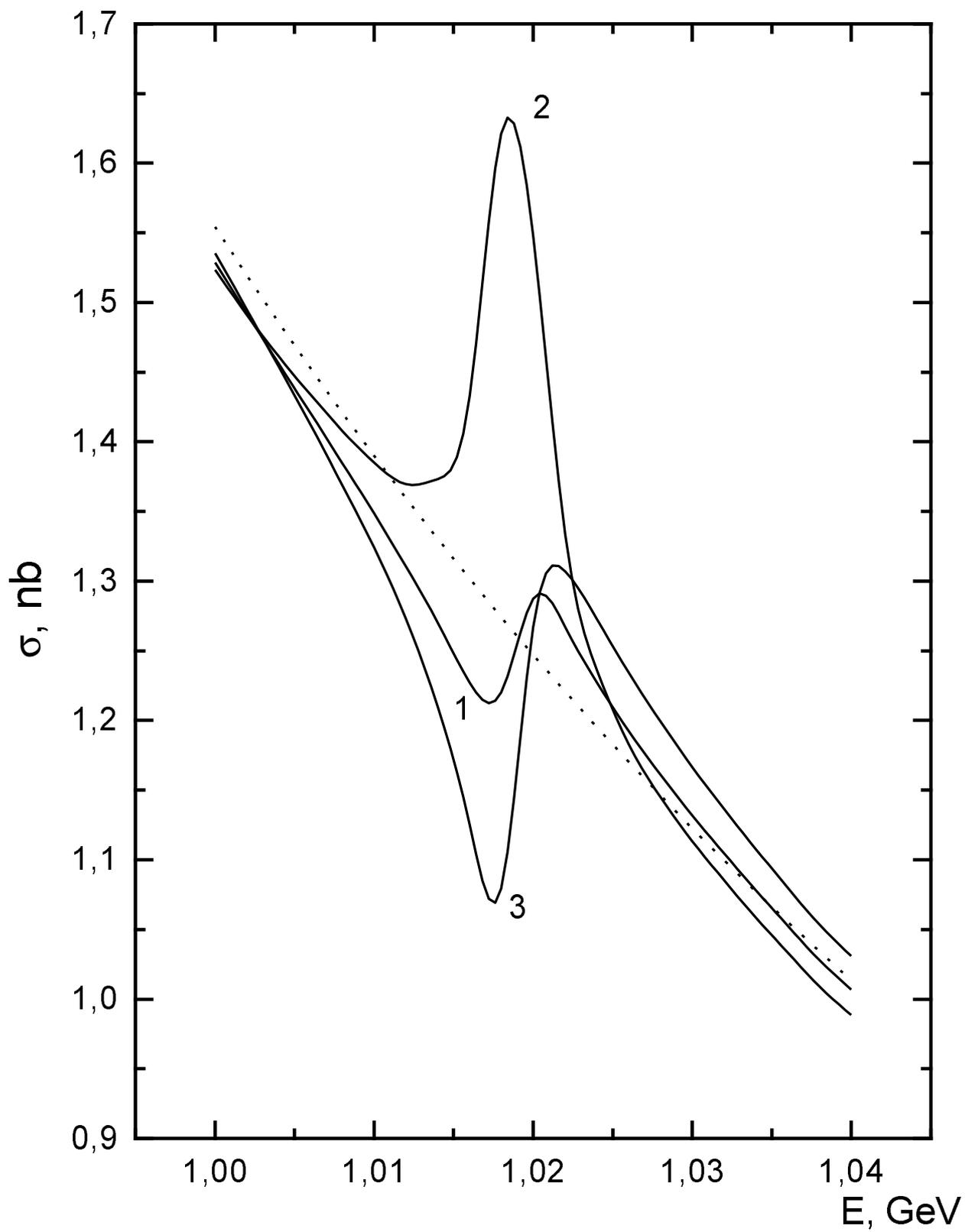
\begin{figure}
\caption{The interference pattern in the total cross section:
$\sigma=\sigma_{\phi}\pm\sigma_{int}+\sigma_f+\sigma_i$ for the $q^2\bar q^2$
 model.
$R=8$,  $g^2_{f_0K^+K^-}/4\pi=2,3\ GeV^2$,
$BR(\phi\to\gamma f_0\to\gamma\pi\pi)=2,3\cdot10^{-4}$,
$BR(\phi\to\gamma f_0\to\gamma\pi\pi,20<\omega<100\ MeV)=1,3\cdot10^{-4}$,
the visible width is $25\ MeV$.
Dotted line is the pure background, the line 1 is the background with 
the  $\phi-\rho$ transition, the line 2 is the constructive interference,
the line 3 is the destructive one.}
\end{figure}
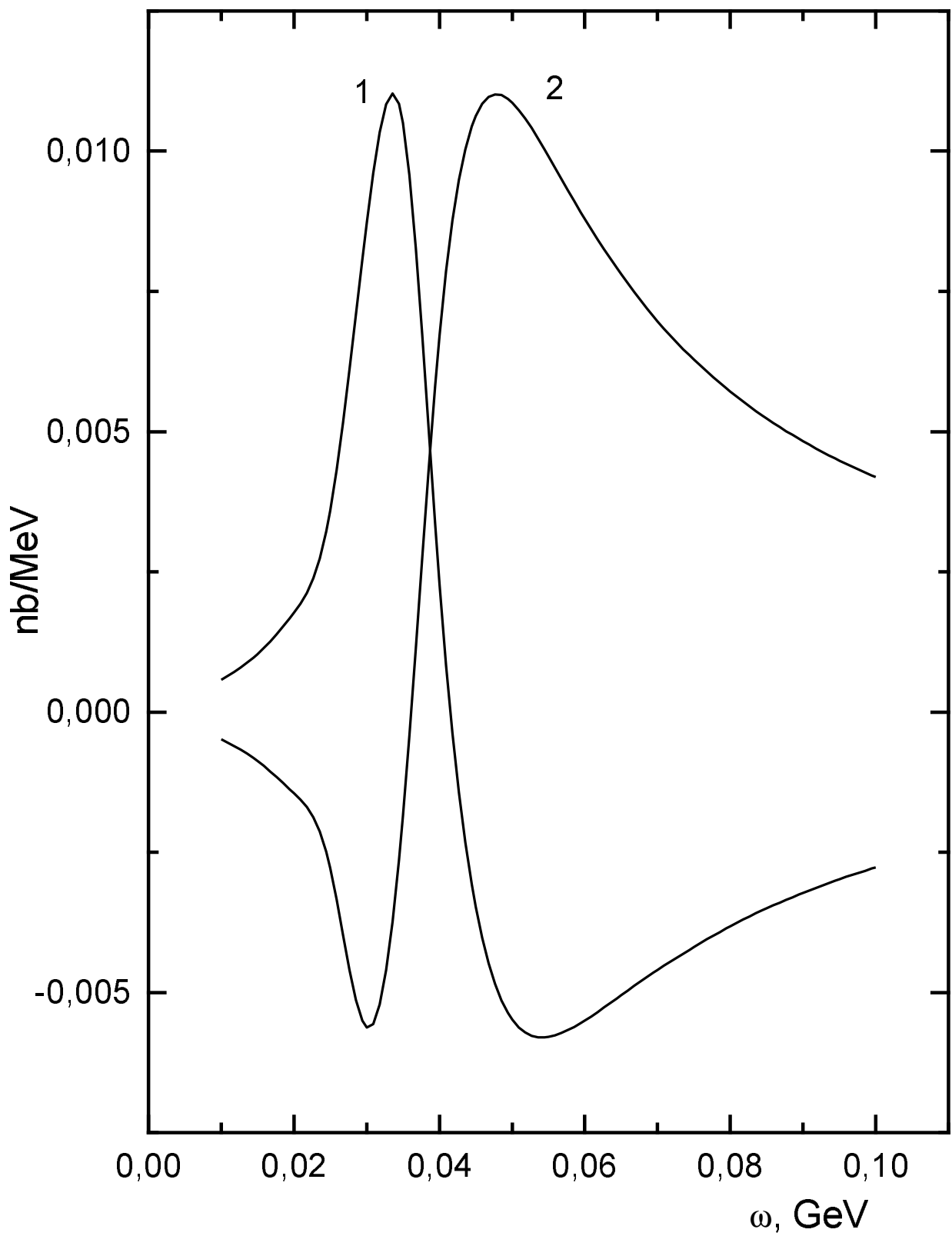
\begin{figure}
\caption{The interference pattern in the photon spectrum:
$d\sigma_{\phi}/d\omega\pm d\sigma_{int}/d\omega$ for the $q^2\bar q^2$
model.
$R=8$,  $g^2_{f_0K^+K^-}/4\pi=2,3\ GeV^2$,
$BR(\phi\to\gamma f_0\to\gamma\pi\pi)=2,3\cdot10^{-4}$.
$BR(\phi\to\gamma f_0\to\gamma\pi\pi,20<\omega<100\ MeV)=1,3\cdot10^{-4}$,
the visible width is $25\ MeV$.
The line 1 is the destructive interference, the line 2 is the constructive
one.}
\end{figure}

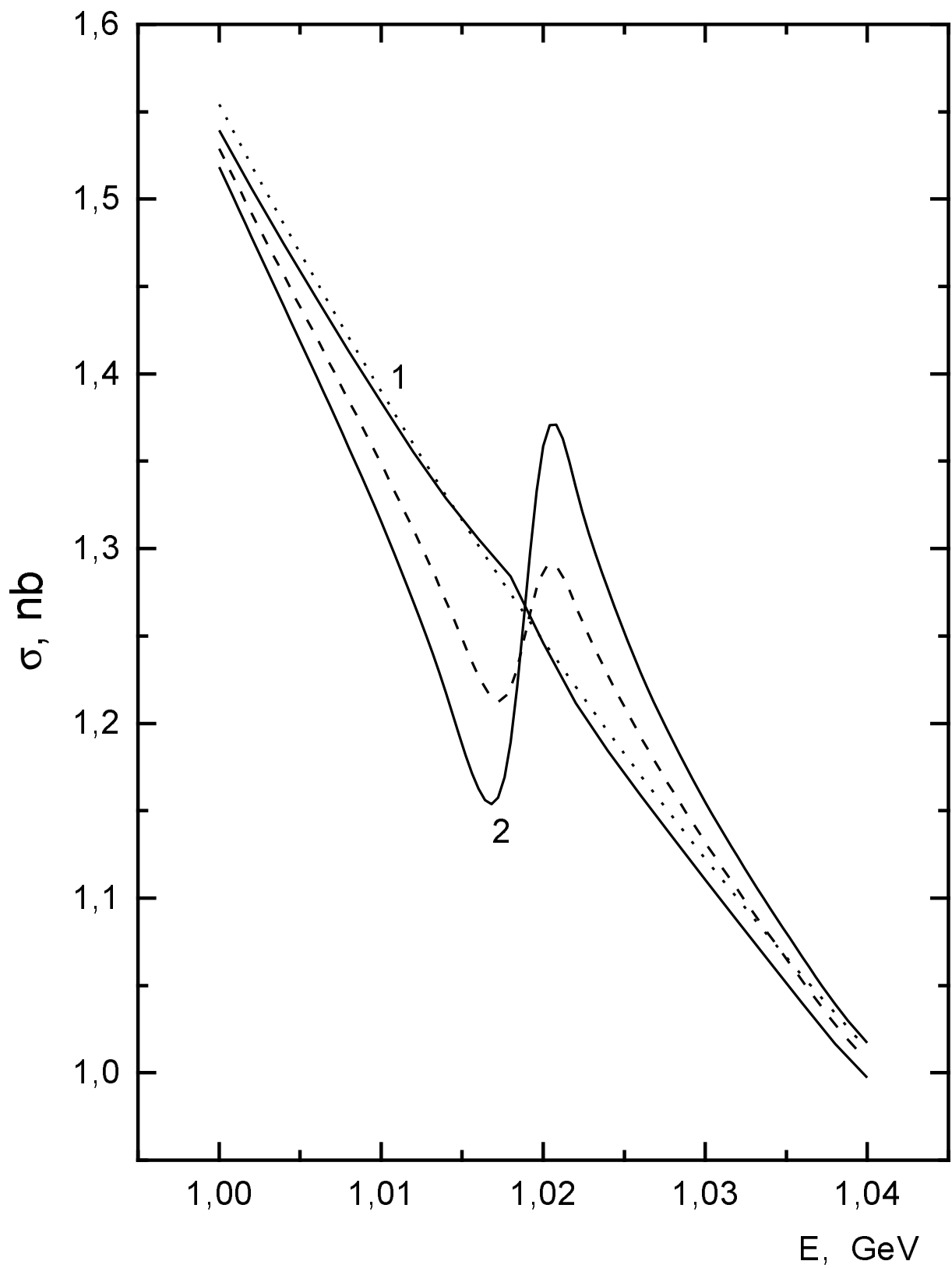
\begin{figure}
\caption{The interference pattern in the total cross section for the model of
the $K\bar K$ molecule.
$\Gamma(f_0\to\pi\pi,m_{f_0}=980\ MeV)=\Gamma_0=50\ MeV$,
the visible width is $25\ MeV$.
$BR(\phi\to\gamma f_0\to\gamma\pi\pi)=1,7\cdot10^{-5}$,
$BR(\phi\to\gamma f_0\to\gamma\pi\pi,20<\omega<100\ MeV)=1,4\cdot10^{-5}$.
The dotted line is the pure background, the dashed line is the background
with the $\phi-\rho$ transition, the line 1 is the constructive
interference, the line 2 is the destructive one.}
\end{figure}

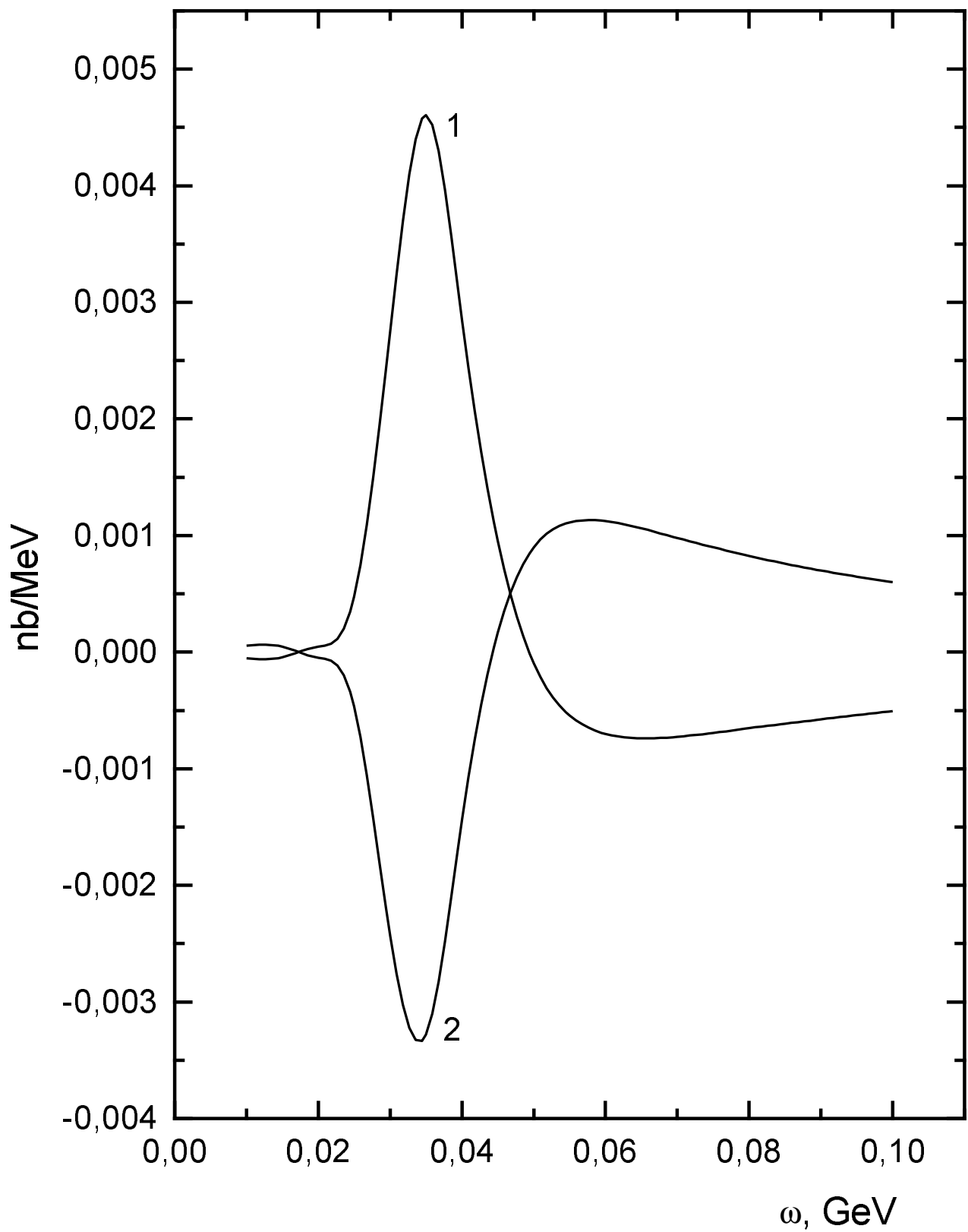
\begin{figure}
\caption{The interference pattern in the photon spectrum for the model of
the $K\bar K$ molecule.
$\Gamma(f_0\to\pi\pi,m_{f_0}=980\ MeV)=\Gamma_0=50\ MeV$,
the visible width is o$25\ MeV$.
$BR(\phi\to\gamma f_0\to\gamma\pi\pi)=1,7\cdot10^{-5}$,
$BR(\phi\to\gamma f_0\to\gamma\pi\pi,20<\omega<100\ MeV)=1,4\cdot10^{-5}$.
The line 1 is the destructive interference, the line 2 is the constructive
one. }
\end{figure}

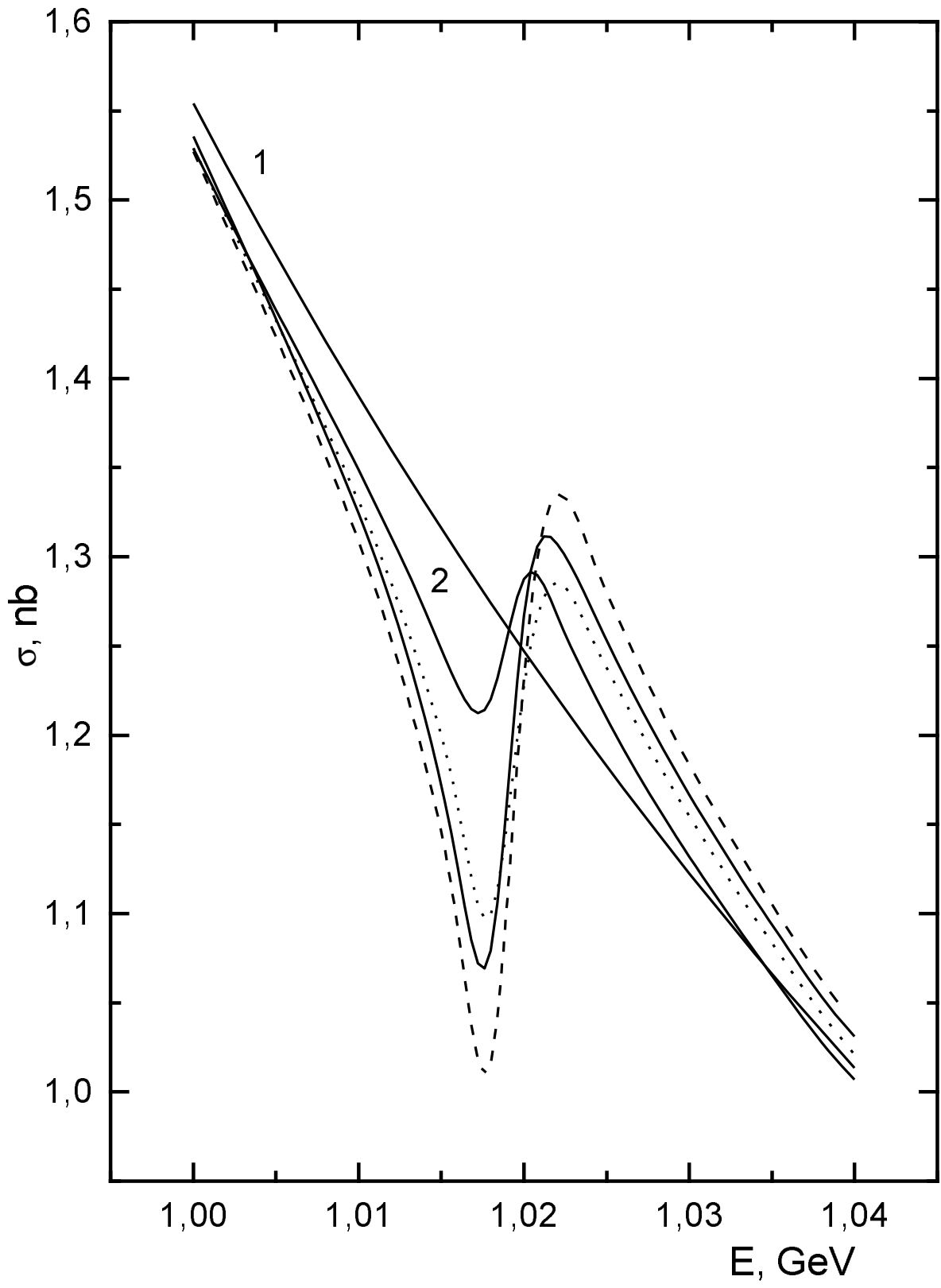
\begin{figure}
\caption{The interference pattern in the total cross section
 at the different parameters.
The line  1 is the pure background. The line 2 is the background with
the $\phi-\rho$ transition,
 $BR(\phi\to\gamma\pi^+\pi^-,20<\omega<100)=3,5\cdot10^{-6}$.
The dotted line is the destructive interference for
$R=4$, $g^2_{f_0K^+K^-}/4\pi=4\ GeV^2$,
$BR(\phi\to\gamma f_0\to\gamma\pi\pi)=5\cdot10^{-4}$,
$BR(\phi\to\gamma f_0\to\gamma\pi\pi,20<\omega<100\ MeV)=1,14\cdot10^{-4}$,
the visible width $\Gamma_{f_0}=50\ MeV$.
The solid line is the destructive interference for
 $BR(\phi\to\gamma f_0\to\gamma\pi\pi)=2,3\cdot10^{-4}$,
$BR(\phi\to\gamma f_0\to\gamma\pi\pi,20<\omega<100\ MeV)=1,13\cdot10^{-4}$,
the visible width $\Gamma_{f_0}=25\ MeV$, see Fig.2.
The dashed line is the destructive interference for  $R=1$,
$g^2_{f_0K^+K^-}/4\pi=0,19\ GeV^2$,
$BR(\phi\to\gamma f_0\to\gamma\pi\pi)=5\cdot10^{-5}$,
$BR(\phi\to\gamma f_0\to\gamma\pi\pi,20<\omega<100\ MeV)=2,4\cdot10^{-5}$,
the visible width  $\Gamma_{f_0}=50\ MeV$. }
\end{figure}
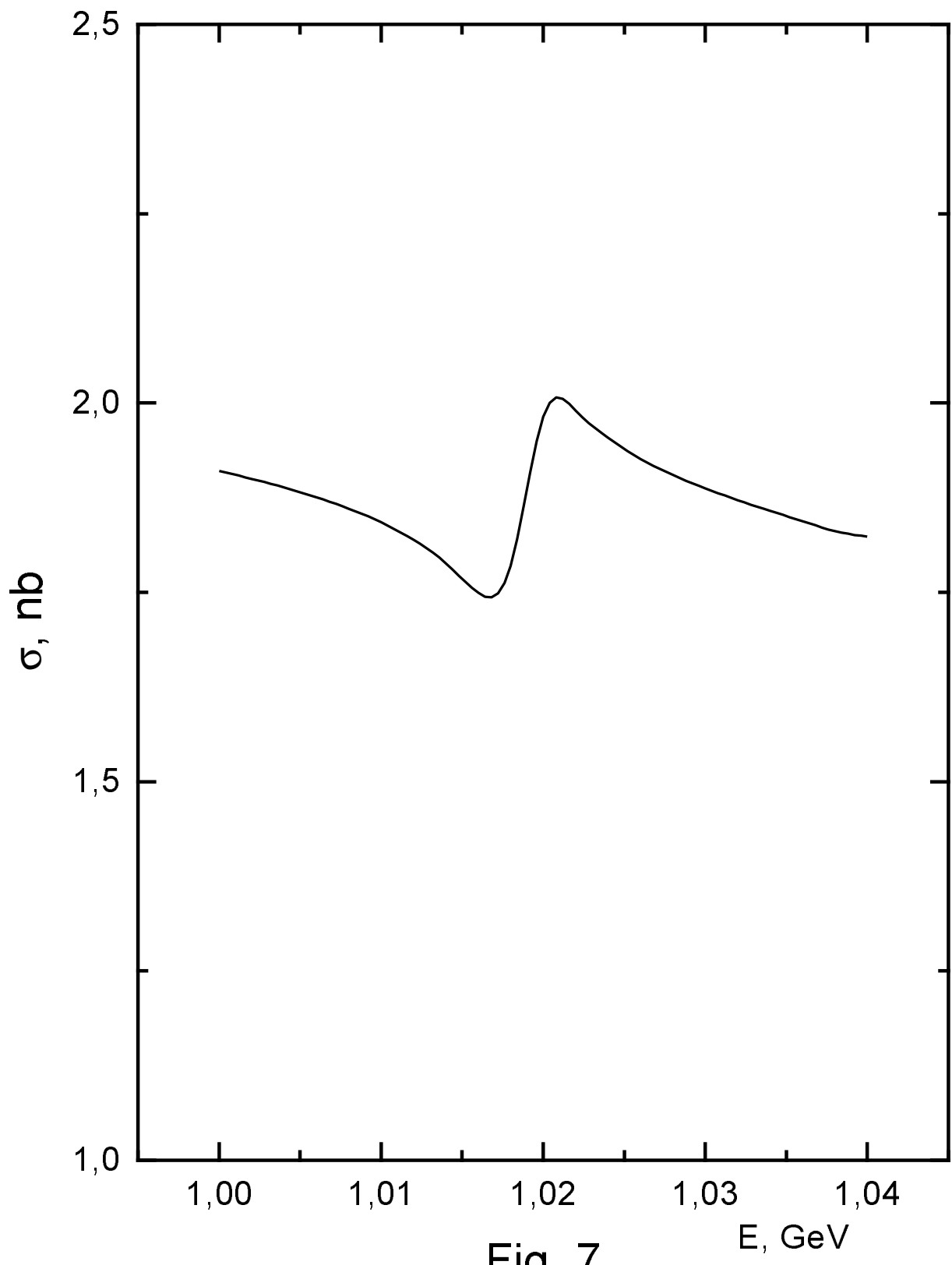
\begin{figure}
\caption{The  $e^+e^-\to\gamma\mu^+\mu^-$ background.
The sum of the initial electron radiation and the final muon one.}
\end{figure}



\begin{references}
\bibitem{achdevshest}
N.N. Achasov, S.A. Devyanin and G.N. Shestakov, Usp. Fiz. Nauk. {\bf142},
361 (1984).
\bibitem{achasov-91}
N.N. Achasov, Nucl. Phys. B (Proc. Suppl.) {\bf21}, 189 (1991).
\bibitem{achshest}
N.N. Achasov and G.N. Shestakov, Usp. Fiz. Nauk {\bf161}, 53 (1991).
\bibitem{jaffe-77}
R.L. Jaffe, Phys. Rev. {\bf D15}, 267, 281 (1977).
\bibitem{weinstein-90}
J. Weinstein and N. Isgur, Phys. Rev. {\bf D41}, 2236 (1990).
\bibitem{tornqvist}
N.A. T$\ddot o$rnqvist Phys. Rev. Lett., {\bf49}, 624 (1982).
\bibitem{braun}
N. Brown and F.E. Close, {\it THE DA$\Phi$NE PHYSICS HANDBOOK}, Vol. II,
 edited by L. Maiani, G. Pancheri, N. Paver, dei Laboratory Nazionali di
 Frascati, Frascati, Italy ( December 1992), p. 447.
\bibitem{gribov}
F.E. Close, Yu.L. Dokshitzer, V.N. Gribov, V.A. Khoze and M.G. Ryskin, \\
Phys. Lett. {\bf B319}, 291 (1993).
\bibitem{achasov-89}
N.N. Achasov and V.N.Ivanchenko, Nucl. Phys. {\bf B315},465 (1989),\\
Preprint INP 87-129 (1987), Novosibirsk.
\bibitem{achasov-95}
N.N. Achasov, {\it THE SECOND DA$\Phi$NE PHYSICS HANDBOOK}, Vol. II, edited by
L. Maiani, G. Pancheri, N. Paver, dei Laboratory Nazionali di Frascati,
Frascati, Italy ( May 1995), p. 671.
\bibitem{close}
F.E. Close, N. Isgur and S. Kumano, Nucl. Phys. {\bf B389}, 513 (1993).
\bibitem{molecule}
N.N. Achasov, V.V. Gubin and V.I. Shevchenko, hep-ph/9605245,
to be published in Yad. Fiz.
\bibitem{novo}
R.R. Akhmetshin et al., preprint Budker INP 95-62, 1995.
\bibitem{snd}
M.N. Achasov et al., preprint Budker INP 96-47, 1996.
\bibitem{bramon}
A.Bramon, G.Colangelo and M.Greco, Phys.Lett. {\bf287}, 263 (1992). \\
A.Bramon et al., {\it The DA$\Phi$NE Physics Handbook}, eds. L.Maiani,
 G.Pancheri \\
and N.Paver, Vol. II, p. 487. \\
A.Bramon, M.Greco, {\it The second DA$\Phi$NE Physics Handbook}, eds.
 L.Maiani, \\
 G.Pancheri and N.Paver, Vol. II, p. 663.
\bibitem{franzini}
J. Lee-Franzini, W.Kim, P.J.Franzini,{\it The DA$\Phi$NE Physics Handbook}, \\
eds. L.Maiani, G.Pancheri and N.Paver, Vol. II, p. 513. \\
J. Lee-Franzini, W.Kim, P.J.Franzini, preprint LNF-92/026 (R),
9 Aprile 1992.\\
G.Colangelo, P.J.Franzini, preprint LNF-92/058 (P), 22 Giugno 1992.
\bibitem{lucio}
J.L.Lucio M., M. Napsuciale, Phys.Lett. {\bf B331}, 418 (1994).
\bibitem{inudequacy}
N.N. Achasov and V.V. Gubin, Phys. Lett. {\bf B 363}, 106 (1995).
\bibitem{bukin}
A.D. Bukin et al., Yad. Fiz. {\bf27}, 985 (1978).
\bibitem{fadin}
E.Kuraev and V.S.Fadin, Yad. Fiz. {\bf41}, 733 (1985).
\bibitem{achasov-88}
N.N. Achasov and G.N. Shestakov, Z. Phys. {\bf C41}, 309 (1988).
\bibitem{xose}
V.N. Baier and V.A. Khose, Zh. Eksp. Teor. Fiz. {\bf48}, 946, 1708 (1965).
\end{references}
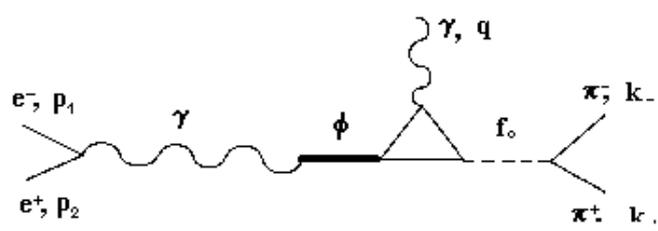
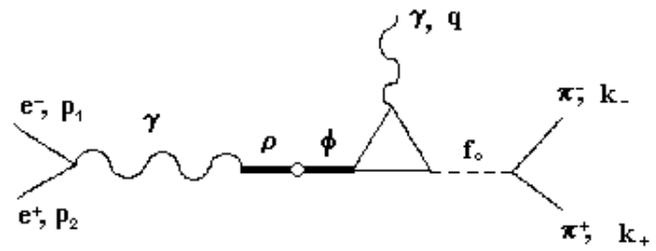
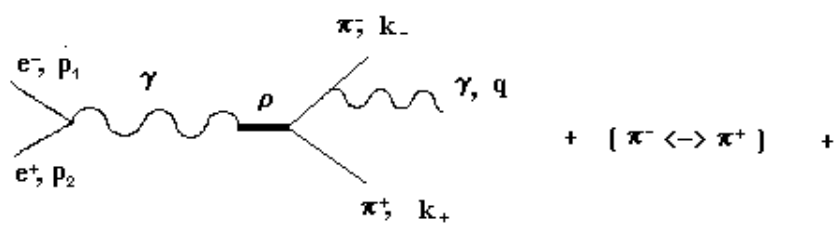
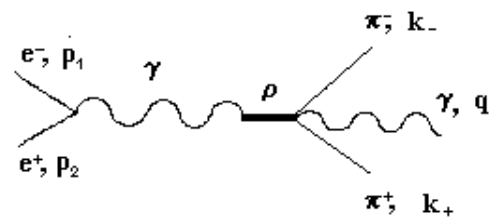
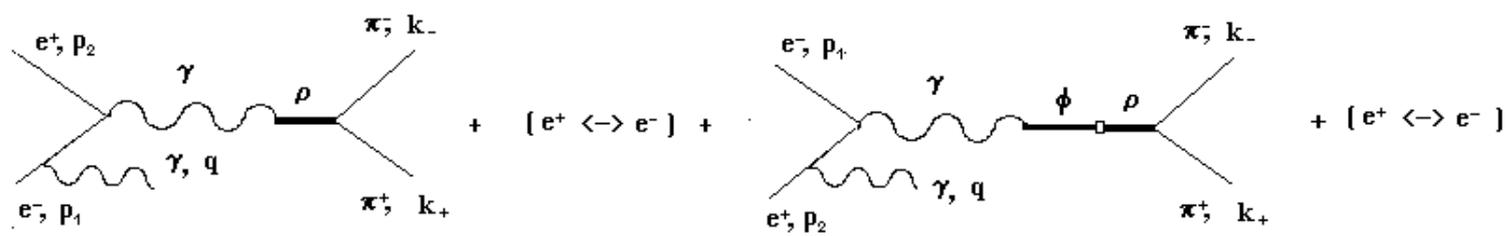
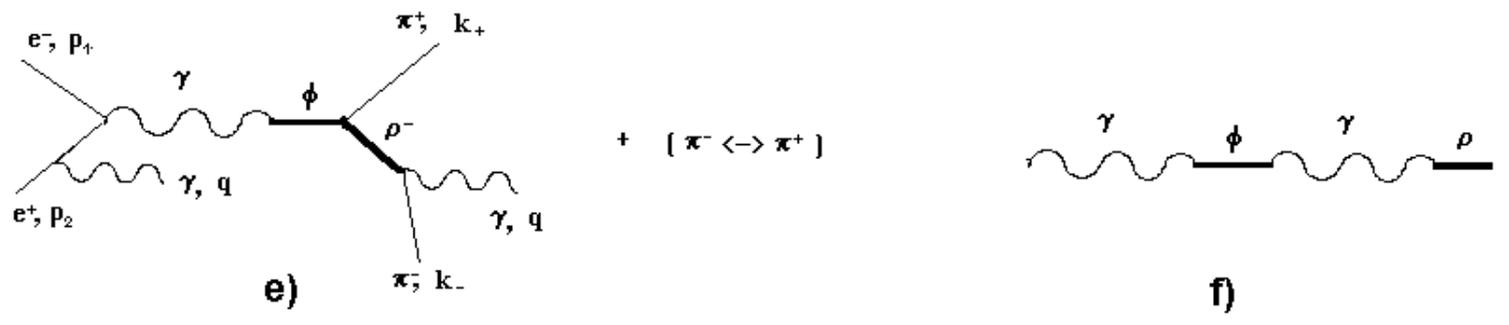
\end{document}